\documentclass[12pt]{iopart}

\pdfoutput=1

\expandafter\let\csname equation*\endcsname\relax
\expandafter\let\csname endequation*\endcsname\relax

\usepackage[english]{babel}
\usepackage{amsmath}
\usepackage{amssymb}
\usepackage{graphicx}
\usepackage{color}
\begin{document}

\title[Large time zero temperature dynamics of the spherical 2-spin glass model of finite size]{Large time zero temperature dynamics of the spherical $p=2$-spin glass model of finite size}

\author{Yan V. Fyodorov}
\address{Queen Mary University of London, School of Mathematical Sciences, London E1 4NS,
United Kingdom}
\ead{y.fyodorov@qmul.ac.uk}

\author{Anthony Perret}
\address{Laboratoire de Physique Th\'eorique et Mod\`eles Statistiques, UMR 8626, Universit\'e Paris Sud 11 and CNRS, B\^at. 100, Orsay F-91405, France}
\ead{anthony.perret@lptms.u-psud.fr}

\author{Gr\'egory Schehr}
\address{Laboratoire de Physique Th\'eorique et Mod\`eles Statistiques, UMR 8626, Universit\'e Paris Sud 11 and CNRS, B\^at. 100, Orsay F-91405, France}
\ead{gregory.schehr@u-psud.fr}

\begin{abstract}
We revisit the long time dynamics of the spherical fully connected $p = 2$-spin glass model when the number of spins $N$ is large but {\it finite}. At $T=0$ where the system is in a (trivial) spin-glass phase, and on long time scale $t \gtrsim {\cal O}{(N^{2/3})}$ we show that the behavior of physical observables, like the energy, correlation and response functions, is controlled by the density of near-extreme eigenvalues at the edge of the spectrum of the coupling matrix $J$, and are thus non self-averaging. We show that the late time decay of these observables, once averaged over the disorder, is controlled by new universal exponents which we compute exactly.
\end{abstract}

\date{\today}

\maketitle


\section{Introduction and model}

Non-equilibrium dynamics of spin-glass models has attracted much interest, both theoretically and
experimentally, during the last 40 years \cite{BB2011}. In particular, the low temperature relaxational dynamics of such systems
following a quench from a high-temperature configuration is not only
extremely slow but also displays ``aging effects''. This means that the response of the system (as well as temporal correlations) depend
strongly on the history of the sample since the temperature quench. It is useful and now customary to characterize quantitatively
such aging effects by studying two-time $t,t'$ observables, including in particular the local response $R(t,t')$ and the auto-correlation function
$C(t,t')$ (see below for a precise definition) \cite{LC2003}.

The analytical computation of these observables for finite-dimensional spin glasses, and more generally for disordered systems in
finite dimension, is a very hard task and consequently there exist very few exact results for such models (see, e.g., Ref. \cite{SLD2004} in the context of disordered elastic systems). However, it was demonstrated
that it is already instructive to study fully-connected (mean-field) spin-glasses, whose dynamical responses and correlations
were shown to exhibit quite rich structures, similar to some extent to the ones observed in structural glasses for instance \cite{BB2011}. The purpose of the present paper is to revisit the late time dynamics of the simplest model of that type, namely the spherical two-spin model of large but finite number of spins, in light of recent results obtained in the literature on random matrix theory (RMT).

The spherical two-spin model, which we focus on here, is defined by the Hamiltonian
\begin{eqnarray}\label{def_SK}
H[\{ s_i\}] = - \frac{1}{2} \sum_{i\neq j} J_{ij} s_i s_j \;,
\end{eqnarray}
where $s_i$, with $i=1,\cdots, N$, are continuous spin variables constrained such that
\begin{eqnarray}\label{spherical}
\sum_{i=1}^N s_i^2 = N
\end{eqnarray}
and where $J$ is a symmetric random matrix belonging to the Gaussian Orthogonal Ensemble (GOE) of random
matrices (corresponding to the Dyson index $\beta =1$): its elements are independently distributed Gaussian random variables with zero mean and variance proportional
to $1/N$ (with this choice, the model (\ref{def_SK}) has a meaningful thermodynamic limit $N \to \infty$). The model was originally introduced in Ref. \cite{KTJ1976} and studied by many authors ever since, see e.g. chap. 4 of the book \cite{DG}. Although it was shown to exhibit a phase transition at the critical temperature $T_c  = 1$ from a paramagnetic phase into a low temperature ``spin-glass'' phase  the latter turns out to be in fact a ``ferromagnetic in disguise''~\cite{DG}. Namely, the system possesses only two ground states related by the symmetry $s_i \to -s_i$ and the calculation of the free energy for this model using replicas does not involve a replica symmetry breaking which is a hallmark of the true spin-glass thermodynamics. As the minimum of the quadratic form (\ref{def_SK}) on a sphere is obviously given by (minus one half of) the largest eigenvalue $\lambda_{\max}$ of the matrix~$J$,
 the statistics of the ground state are governed by the Tracy-Widom distribution for GOE, which describes the fluctuations of the largest/smallest eigenvalue in that ensemble~\cite{TW96}. Besides, it was recently proved that for $T<T_c$, the fluctuations of the free energy are also given by the Tracy-Widom distribution, see  \cite{BL2015}.

Throughout this paper, we will be interested in the limiting case $T=0$, which already contains many interesting aspects. Although, as explained above, the thermodynamics of the model in the low-temperature phase is too simple for a bona fide spin-glass, the corresponding dynamics is rich and has features of aging \cite{cugliandolo_dean,ZKH,BDG2001,BDG2006}. That richness is attributed to a relatively non-trivial energy landscape topology due to the presence of many saddle points in the landscape with different indices $k$ (the number of unstable directions), each associated with the eigenvalues $\lambda_i<\lambda_{\max}$ of the matrix $J$. Note that the presence of a magnetic field in the system leads to a simplification of the associated energy landscape, which gradually washes out the complexity of the relaxational dynamics (see Refs. \cite{CD95,FLD2014,DZ2015,GT2015} for this and related questions). Our studies of the dynamics of the model (\ref{def_SK}) will be intimately related to the setting of the original work by Cugliandolo and Dean \cite{cugliandolo_dean} referred to as CD in the following. In particular, we will use  the same convention for the distribution of the matrix elements $J_{ij}$ such that when $N \to \infty$, the distribution of the eigenvalues of the random matrix $J$ is given by the Wigner semi-circle law on the bounded interval~$[-2,2]$:
\begin{eqnarray}\label{Wigner}
\rho(\lambda) = \frac{1}{2 \pi} \sqrt{4 - \lambda^2} \;.
\end{eqnarray}

To study the relaxational dynamics of this model (\ref{def_SK}), it is convenient to diagonalize the coupling matrix $J$
and to work with the time dependent projections of the spin configuration $s(t) = \{ s_i(t)\}_{1 \leq i \leq N}$ onto the eigenvectors of $J$, which
are denoted by $s_\lambda(t)$, $\lambda$ belonging to the spectrum of $J$, $\lambda \in {\rm Sp}{(J)}$. The dynamics of the model is then defined via a Langevin equation, which when projected onto the
eigenvectors of $J$, yields a Langevin equation for the projections $s_{\lambda}(t)$ that reads \footnote{note that Eq. (\ref{def_Langevin}) also describes the dynamics of a spherical height model, the Arcetri model recently introduced in \cite{HD2015}, in one spatial dimension.}
\begin{eqnarray}\label{def_Langevin}
\frac{\partial s_\lambda(t)}{\partial t} = (\lambda - z(t)) s_\lambda(t) + h_\lambda(t) + \xi_\lambda(t) \;,
\end{eqnarray}
where $h_\lambda(t)$ represents an (infinitesimal) external magnetic field -- which is used here to compute the response
function -- and $z(t)$ is a Lagrange multiplier which enforces the spherical constraint (\ref{spherical}). In Eq. (\ref{def_Langevin}), $\xi_\lambda(t)$ is a Gaussian
white noise of zero mean and correlations
\begin{eqnarray}
\langle \xi_\lambda(t) \xi_{\lambda'}(t') \rangle = 2 T \delta_{\lambda, \lambda'} \delta(t-t') \;,
\end{eqnarray}
where $T$ is the temperature and where $\langle \cdots \rangle$ denotes an average over the thermal noise.

\section{Summary of main results of the paper}

Here we consider the zero temperature dynamics, at $T=0$, where the system is quenched at initial time $t=0$ from a high
temperature configuration, described by a uniform initial condition
\begin{eqnarray}\label{CI}
s_\lambda(0) = 1 \;, \; \forall \lambda \in {\rm Sp}(J) \;,
\end{eqnarray}
which was shown to lead to a rich dynamical behavior \cite{cugliandolo_dean}. In the large time limit, it is easy to see that the Langevin dynamics (\ref{def_Langevin}) at $T=0$, i.e. $\xi_\lambda(t) = 0$ (assuming $h_\lambda(t)=0$), will drive the system to relax to the
configuration with minimal energy per spin (energy density) $e(t \to \infty) = -\frac{1}{2} \lambda_{\max}$ where
\begin{eqnarray}
\lambda_{\max} = \max_{\lambda \in {\rm Sp}\,{(J)} } \lambda \;,
\end{eqnarray}
 and correspondingly $s_{\lambda}(t \to \infty) \to \sqrt{N} \delta_{\lambda,\lambda_{\max}}$. The question that we will address in the present paper is the following: how does the system approach this final state of minimal energy? For instance, how does the average energy density $\overline{e}(t)$ approach its limiting value?
To characterize the non-equilibrium dynamics of the system (\ref{def_Langevin}) it is also useful to consider quantities depending on two times $t,t'$. Here we will focus on the zero-temperature limit of the spin-spin correlation function $C(t,t')$ and on the response function $R(t,t')$, $t>t'$:
\begin{eqnarray}\label{def_CR}
&&{C(t,t')} = \frac{1}{N} \sum_{i=1}^N {s_i(t) s_i(t')} = \frac{1}{N}\sum_{\lambda} {s_\lambda(t) s_{\lambda}(t')} \;, \\
&&{R(t,t')} =\frac{1}{N} \sum_{i=1}^N{\frac{\delta s_i(t) }{\delta h_i(t')}\Bigg |_{h=0}} =\frac{1}{N}{\sum_{\lambda} \frac{\delta s_\lambda(t) }{\delta h_\lambda(t')} \Bigg |_{h=0}} \nonumber \;,
\end{eqnarray}
and their corresponding disorder averaged values $\overline{C(t,t')}$ and $\overline{R(t,t')}$ where $\overline{\cdots}$ denotes an average over the disorder realizations, i.e. over the GOE random coupling matrix $J$. In Eq. (\ref{def_CR}), and in the following, `$\sum_{\lambda}$' denotes a sum over all the eigenvalues belonging to the spectrum of $J$.

Intuitively, one expects that at large time, the dynamics in Eq. (\ref{def_Langevin}) will be dominated by the near-extreme eigenvalues, i.e. the eigenvalues of the coupling matrix $J$ that are close to $\lambda_{\max}$ \cite{cugliandolo_dean,KL1996}. More precisely, we will see that the observables mentioned above (energy density, response and correlation functions) can be written in terms of the density of eigenvalues ``seen'' from $\lambda_{\max}$, the so called density of states (DOS) $\rho_{\rm DOS}(r,N)$ defined as \cite{SM2007,perret_schehr}
\begin{eqnarray}\label{def_DOS}
\rho_{\rm DOS}(r,N) = \frac{1}{N-1} \sum_{\lambda \neq \lambda_{\max}} \delta(\lambda_{\max} - \lambda - r) \;.
\end{eqnarray}
The DOS (\ref{def_DOS}) was recently studied in detail for matrices belonging to the Gaussian Unitary Ensemble (GUE, corresponding to the Dyson index $\beta = 2$) \cite{perret_schehr}, using semi-classical orthogonal polynomials, as well as for more general Gaussian $\beta$-ensembles \cite{perret_schehr2}, using mainly scaling arguments. In particular, it was shown that the behavior of the average DOS $\overline{\rho_{\rm DOS}(r,N)}$ exhibits two distinct behaviors depending on whether $r \sim {\cal O}(1)$, or $r \sim {\cal O}(N^{-2/3})$. In the former case $\overline{\rho_{\rm DOS}(r,N)}$ is simply given by a shifted Wigner semi-circle law (see also \cite{KL1996}), whereas
 in the latter $\overline{\rho_{\rm DOS}(r,N)}$ takes a non-trivial scaling form reflecting  the fluctuations at the edge of the Wigner semi-circle. This can be summarized as follows \cite{perret_schehr,perret_schehr2}
\begin{eqnarray}\label{summary}
\overline{\rho_{\rm DOS}(r,N)} \sim
\begin{cases}
&\tilde \rho_{\rm bulk}(r) = \frac{1}{2 \pi} \sqrt{r(4-r)} \;, \; r \sim {\cal O}(1) \;,\\
& \\
&N^{-1/3}\tilde \rho_{\rm edge}(N^{2/3} r) \;, \; r \sim {\cal O}(N^{-2/3}) \;.
\end{cases}
\end{eqnarray}
The scaling function $\tilde \rho_{\rm edge}(x)$ is presently known exactly only for complex Hermitian random matrices belonging to GUE \cite{perret_schehr}. For general Gaussian $\beta$-ensembles, including the case $\beta =1$ mostly relevant for the spin-glass model studied here (\ref{def_SK}), the exact evaluation of $\tilde \rho_{\rm edge}(x)$ remains an open problem. At the same time the asymptotic behaviors of such a scaling function was derived in \cite{perret_schehr2} and is given by
\begin{eqnarray}\label{asympt_redge}
\tilde \rho_{\rm edge}(x) \sim
\begin{cases}
&a_\beta \, x^\beta \;, \; x \to 0 \;, \\
& \\
&\frac{1}{\pi} \sqrt{x} \;,  x \to \infty  \;.
\end{cases}
\end{eqnarray}
The small-$x$ asymptotics in (\ref{asympt_redge}) is controlled by the repulsion between two neighboring eigenvalues, and the exact value of the amplitude $a_\beta$ is known only for $\beta = 2$ where $a_2 = 1/2$. On the other hand,
the large argument behavior can be obtained by matching the edge regime with the bulk regime given by the shifted Wigner semi-circle.

In Ref. \cite{cugliandolo_dean} (see also Ref. \cite{BDG2001,BDG2006} for a rigorous proof of these results), the authors studied the dynamics of the spherical $2$-spin model in the thermodynamic limit $N \to \infty$. Most importantly the limit of large times $t$ was studied {\it after} performing the limit $N \to \infty$. In this order of limits the large time behavior of physical observables can be obtained by replacing $\rho_{\rm DOS}(r,N)$ (which is self-averaging for $r \sim {\cal O}(1)$) by a shifted Wigner semi-circle $\tilde \rho_{\rm bulk}(r)$ given by the first line of (\ref{summary}). Indeed, in the limit $N \to \infty$ the contribution from the edge regime turns out to be negligible. In particular, it was shown that the average energy density $\overline{e}(t)$  approaches its limiting value algebraically as
\begin{eqnarray}\label{energy_CD}
\overline{e}(t) + \frac{1}{2} \overline{\lambda_{\max}} \sim \frac{3}{8\,t} \;,
\end{eqnarray}
for large time $t$. Such algebraic decay reflects the complexity of the energy landscape, with infinitely many saddle-points  being operative in trapping the descending trajectories for a long time. Such saddle points typically
have a ``mesoscopic'' index, that is the number $k$ of unstable directions satisfy $1\ll k\ll N$, corresponding to eigenvalues of the matrix $J$  mesoscopically far from the spectral edge. We notice that a similar decay $\propto 1/t$, as in Eq. (\ref{energy_CD}), was also observed in numerical
simulations of the Sherrington-Kirkpatrick model \cite{deltae_SK} (at $T=T_c$).

 In this paper, we are interested in clarifying what happens for such a system of large but finite size $N$. In this case, we actually demonstrate that there exists a crossover time scale
$t_{\rm cross} \sim {\cal O}(N^{2/3})$ which separates two distinct relaxation regimes: the CD regime 1 as described above for the system in the thermodynamic limit $N\to \infty$ and taking place for times $t \ll t_{\rm cross}$  from the later time regime 2 operative for $t \gg t_{\rm cross}$. This latter regime is controlled by the small argument behavior of $\tilde \rho_{\rm edge}(x)$ in Eq. (\ref{asympt_redge}), yielding in particular new exponents which we compute here. At this later stage the relaxation is dominated by just a few saddle-points in the landscape close in energy to the ground state and having of order of one unstable directions.

\begin{figure}
\centering
\includegraphics[width = 0.8\linewidth]{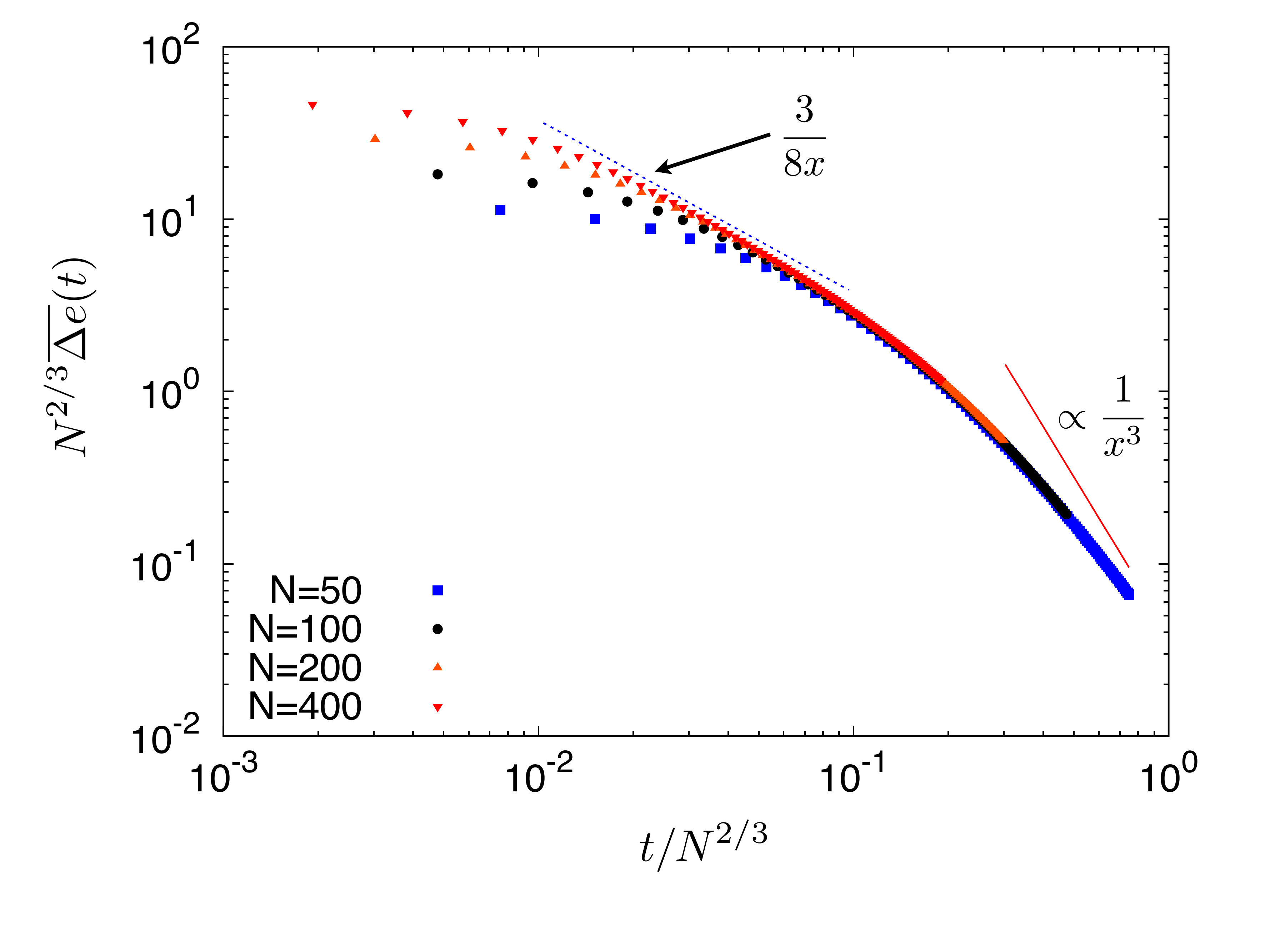}
\caption{Plot of $N^{2/3} \overline{\Delta e}(t)$ as a function of $t/N^{2/3}$ for different values of $N=50,100,200$ and $400$. The data have been obtained by evaluating numerically the exact formula in Eqs. (\ref{expr_z_2})  and (\ref{explicit_e}), and the averages have been performed by sampling 200 independent GOE random matrices. The collapse of the curves for different values of $N$ on a single master curve (for $t$ sufficiently large) is in good agreement with our analytical predictions in Eqs.~(\ref{deltae_intro}) and (\ref{asympt_E}).}\label{fig_energy}
\end{figure}

For $t \sim {\cal O}(N^{2/3})$, single time quantities like the average energy $\overline{e}(t)$ are characterized by a scaling function of the scaling variable $t/N^{2/3}$ which interpolates between these two regimes. For instance, we show that average excess energy $\overline{\Delta e}(t) = \overline{e}(t) +  \overline{\lambda_{\max}}/2$ behaves as
\begin{eqnarray}\label{deltae_intro}
\overline{\Delta e}(t) \sim
\begin{cases}
&e_1(t) \;, \; t \ll N^{2/3} \;, \\
&\\
& N^{-2/3} {\cal E}\left( \dfrac{t}{N^{2/3}}\right) \;, \; t \gtrsim {\cal O}(N^{2/3})\;.
\end{cases}
\end{eqnarray}
The function $e_1(t)$ was computed by CD in Ref. \cite{cugliandolo_dean} and has the asymptotic behavior, for large $t$, given in Eq. (\ref{energy_CD}) while ${\cal E}(x)$ has the following asymptotic behaviors
\begin{eqnarray}\label{asympt_E}
{\cal E}(x) \sim
\begin{cases}
&\dfrac{3}{8 x} \;, \; x \to 0 \;, \\
& \\
&  \dfrac{A}{x^{3}}  \;, \; x \to \infty \;,
\end{cases}
\end{eqnarray}
where $A$ is an (unknown) numerical constant. The regime $x \to 0$ naturally matches the CD regime in Eq. (\ref{energy_CD}), while the other regime for $x \to \infty$, characterized by a different exponent, describes the late time behavior for a system of large but finite size. In Fig.~\ref{fig_energy} we show a plot of $N^{2/3} \overline{\Delta e}(t)$, evaluated numerically from the exact formula derived in Eqs. (\ref{expr_z_2}) and (\ref{explicit_e}) below, as a function of $t/N^{2/3}$ for different values of $N$, which confirms our analytical predictions in Eqs.~(\ref{deltae_intro}) and (\ref{asympt_E}). It is important to stress that while in the regime 1, for $t \ll N^{2/3}$ the quantities we study are self-averaging, we will argue that it is not the case for the late regime 2, for $t \gtrsim N^{2/3}$.

\begin{figure}
\centering
\includegraphics[width = 0.7\linewidth]{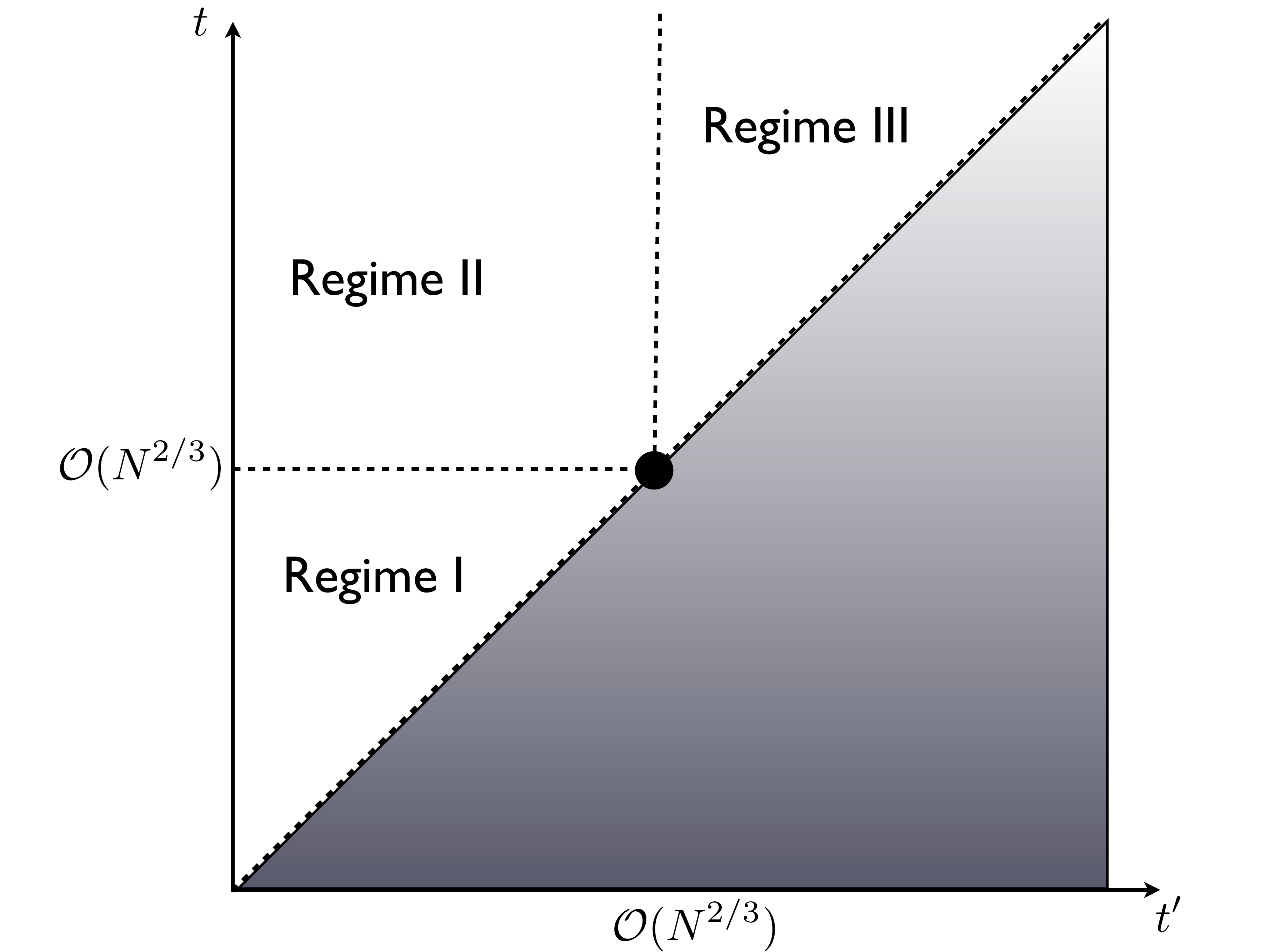}
\caption{The three different regimes of the temporal behavior of the response function $\overline{R(t,t')}$ in the $(t,t')$ plane (note that the region $t<t'$ is forbidden as a consequence of causality). The regime I corresponds to the regime studied by CD in \cite{cugliandolo_dean} while the two regimes had not been studied before. In particular, in the regime III where the dynamics is stationary, the response function is completely determined by the density of near extreme eigenvalues of the coupling matrix $J$ [see Eq. (\ref{rel_r_DOS}) ]. The dot indicates the region where the response function takes the scaling form given in Eq. (\ref{crossover_regime}) which interpolates between the three regimes.}\label{Fig:ph_diag}
\end{figure}

Similarly, two-time observables like the response $\overline{R(t,t')}$ and the correlation $\overline{C(t,t')}$ display different asymptotic behaviors in the three regions I, II and III in the $(t,t')$ plane depicted in Fig. \ref{Fig:ph_diag}. The first region I for times $1 \ll t,t'\ll N^{2/3}$  studied by CD \cite{cugliandolo_dean}, the region II for $1 \ll t' \ll N^{2/3} \ll  t$  and finally the region III for $N^{2/3} \ll t,t'$. In particular, in the region III the response function is stationary, i.e. it only depends on the time difference $t-t'$, and $\overline{R(t,t')}$ is given by
\begin{eqnarray}\label{rel_r_DOS}
\hspace*{-1.cm}\overline{R(t,t')} \sim R_{\rm III}(t-t') = \int_{0}^\infty \overline{\rho_{\rm DOS}(r)} \e^{-r(t-t')}\, dr\;,
\end{eqnarray}
which shows the physical relevance of the DOS in this problem. The scaling form for the DOS in Eq. (\ref{summary}) induces the following scaling form
for the response function $R_{\rm III}(\tau)$
\begin{eqnarray}\label{summary_rIII}
R_{\rm III}(\tau) \sim
\begin{cases}
&r_1(\tau) \;, \; \tau \ll N^{2/3} \\
& \\
& N^{-1} \, \tilde r\left( \dfrac{\tau}{N^{2/3}}\right) \;, \; \tau \gtrsim N^{2/3} \;,
\end{cases}
\end{eqnarray}
where the function $r_1(\tau)$ was computed by CD \cite{cugliandolo_dean} while $\tilde r(x)$ is given by
\begin{eqnarray}\label{rtilde}
\tilde r(x) = \int_0^\infty \e^{-x r} \tilde \rho_{\rm edge}(r) \, dr \;.
\end{eqnarray}
Its small and large argument asymptotic can be easily obtained from the ones for $\tilde \rho_{\rm edge}(r)$ in Eq. (\ref{summary}) with $\beta = 1$ and are given by
\begin{eqnarray}
\tilde r(x) \sim
\begin{cases}
&\dfrac{1}{4\sqrt{2\pi}}\,\dfrac{1}{x^{3/2}} \;, \; x \to 0 \;,\\
& \\
&\dfrac{a_1}{x^{2}}\;, \; x \to \infty \;,
\end{cases}
\end{eqnarray}
where $a_1$ is the amplitude in Eq. (\ref{asympt_redge}) for $\beta = 1$. The limit $x \to 0$ naturally matches the result of CD while the large-$x$ behavior gives rise to a different algebraic decay. This late time regime III corresponds to the final stage of relaxation within the multidimensional energy landscape dominated by both the global minimum and the small-index saddle-points.

Finally, for $t\sim N^{2/3}$ and $t'\sim N^{2/3}$, the response and the correlation functions are characterized by scaling functions of the two variables $t/N^{2/3}$ and $t'/N^{2/3}$ whose asymptotic behaviors match smoothly with the various regimes described above. For instance, the response function $\overline{R(t,t')}$ takes the scaling form
\begin{eqnarray}\label{crossover_regime}
\overline{R(t,t')} = \frac{1}{N} {\cal R}\left(\tilde t = \frac{t}{N^{2/3}}, \tilde t'=\frac{t'}{N^{2/3}} \right) \;, \; t>t' \;,
\end{eqnarray}
which is in general not stationary. Such a non-stationarity is usually interpreted as a form of `aging' typical for complex systems \cite{cugliandolo_dean}. However, in the limit $\tilde t, \tilde t' \gg 1$ stationarity is restored so that ${\cal R}(\tilde t,\tilde t') \sim \tilde r(\tilde t - \tilde t')$, where $\tilde r(x)$ is given in Eq. (\ref{rtilde}).

\section{Zero temperature dynamics}

The general solution of (\ref{def_Langevin}), starting from a given initial condition $s_\lambda(0)$ at $t=0$ for $\lambda \in {\rm Sp}(J)$ was found by CD and reads~\cite{cugliandolo_dean}:
\begin{eqnarray}\label{solution}
&&s_\lambda(t) = s_\lambda(0) \exp(\lambda t) \exp{\left[-\int_0^t z(\tau) d\tau \right]} \\ \nonumber
&&+ \int_0^t dt'' \exp{[\lambda(t-t'')]} \exp{\left[- \int_{t''}^t z(\tau') d\tau' \right]\left(h_\lambda(t'') + \xi_\lambda(t'')\right)} \;,
\end{eqnarray}
which will be the starting point of our analytical computations. In the following, we will focus on the uniform initial condition $s_\lambda(0) =1$, see Eq. (\ref{CI}).

\subsection{Lagrange multiplier $z(t)$ and energy density $e(t)$}

At $T=0$, the expression of $z(t)$ is simply obtained from the spherical condition ${C(t,t)} =1$. Using the definition of ${C(t,t)}$ in Eq. (\ref{def_CR}) together with the exact solution of the Langevin dynamics in Eq. (\ref{solution}), then
setting $h_\lambda(t') = \xi_\lambda(t') = 0$ for all $0<t' \leq t$ (and all $\lambda$'s) one obtains:
\begin{eqnarray}\label{spherical_condition}
N = \sum_{\lambda} \exp{\left[-2 \int_0^t (z(\tau)-\lambda) d\tau\right]} \;.
\end{eqnarray}
After simple algebra the spherical condition (\ref{spherical_condition}) can be represented as:
\begin{eqnarray}\label{expr_z_1}
z(t) = \frac{1}{2} \frac{d}{dt} \ln{\left(\frac{1}{N} \sum_{\lambda} \exp{(2 \lambda t)} \right)} \;,
\end{eqnarray}
which by separating the contribution from the maximal eigenvalue $\lambda_{\max}$ can be conveniently rewritten in the form
\begin{eqnarray}\label{expr_z_2}
&& z(t) = \lambda_{\max} + \frac{1}{2} \frac{d}{dt} \ln g_N(t) \;, \; \\
&& g_N(t) = \frac{1}{N} \sum_\lambda \e^{2(\lambda-\lambda_{\max}) t} = \frac{1}{N} + \frac{1}{N}\sum_{\lambda \neq \lambda_{\max}} \e^{2 (\lambda-\lambda_{\max})t} \;. \nonumber
\end{eqnarray}
From Eq.~(\ref{expr_z_2}) it is obvious that
\begin{eqnarray}
\lim_{t \to \infty} z(t) = \lambda_{\max} \;.
\end{eqnarray}
The finite time behavior of $z(t)$ is therefore controlled by the random function $g_N(t)$ in~(\ref{expr_z_2}). It is useful to write the latter function in terms of the density of near-extreme eigenvalues given in Eq. (\ref{def_DOS}) as
\begin{eqnarray}\label{def_hN}
g_N(t) = \frac{1}{N} + \frac{N-1}{N} h_N(t) \;, \; h_N(t) = \int_0^\infty \e^{-2 r t} \rho_{\rm DOS}(r,N) dr \;.
\end{eqnarray}
From the results for $\overline{\rho_{\rm DOS}(r,N)}$ \cite{perret_schehr2} stated above in Eq. (\ref{summary}), we expect a very different behavior of  $h_N(t)$ for $t \sim {\cal O}(1)$ (and more generally for times of order of $t \ll N^{2/3})$, and for longer times $t \gtrsim {\cal O}(N^{2/3})$.

Indeed, for $t \sim {\cal O}(1)$, the integral in Eq. (\ref{def_hN}) is dominated by the region $r \sim {\cal O}(1)$ where the DOS is (i) self-averaging and (ii) given by the shifted Wigner semi-circle law $\tilde \rho_{\rm bulk}(r)$ in Eq. (\ref{summary}). Therefore for this range of times we have:
\begin{eqnarray}
\label{self_av_order1}
\lim_{N \to \infty} h_N(t) = \overline{h_N(t)} = \bar{h}(t) = \int_0^4 \e^{-2 r t} \tilde \rho_{\rm bulk}(r) \, dr, \; {\rm for} \; t \sim {\cal O}(1) \;.
\end{eqnarray}
The integral over $r$ can be performed explicitly \cite{cugliandolo_dean} which yields \footnote{Note that $\bar h(t) = \e^{-4t} \Gamma(t)$ in the notation of Ref. \cite{cugliandolo_dean}.}
\begin{eqnarray}\label{self_av_order2}
 \bar{h}(t) = \int_0^4 \frac{dr}{2\pi} \, \e^{-2 r t} \sqrt{r(4-r)} = \frac{\e^{-4t}}{2t} I_1(4t) \;,
\end{eqnarray}
where $I_\nu(x)$ is the modified Bessel function of index $\nu$.  Eq. (\ref{self_av_order2}) implies that $\bar h(t)$ is a monotonically decreasing function of $t$ whose asymptotic behaviors are given by
\begin{eqnarray}
\bar h(t) \sim
\begin{cases}
& 1 - 4 t \;, \; t \to 0 \;, \\
& \\
&\dfrac{1}{4 \sqrt{2 \pi}} t^{-3/2} \;, \; t \gg 1 \;.
\end{cases}
\end{eqnarray}

On the other hand, for $t \sim {\cal O}(N^{2/3})$ one expects that the integral in Eq. (\ref{def_hN}) will be dominated by $r \sim {\cal O}(N^{-2/3})$ where the DOS behaves differently, see Eqs. (\ref{summary})~and~(\ref{asympt_redge}). For such values of $r$ one expects that the DOS is not a self-averaging quantity, which implies that $h_N(t)$, for $t \sim {\cal O}(N^{2/3})$ is not self averaging either -- and we have checked this statement numerically in \ref{appendix_self_av}. Finally, for very large times , $t \gg {\cal O}(N^{2/3})$, the function $h_N(t)$ is obviously dominated by the first gap between the two largest eigenvalues $g = \lambda_1 - \lambda_2$, where we have used the ordering $\lambda_{\max} = \lambda_1 > \lambda_2 > \cdots > \lambda_N$. We then arrive at the following asymptotic behaviors for $h_N(t)$:
\begin{eqnarray}\label{asympt_hN}
h_N(t) \sim
\begin{cases}
&\dfrac{1}{4 \sqrt{2 \pi}} t^{-3/2} \;, \; 1 \ll t \ll N^{2/3} \;, \\
& \\
&\dfrac{1}{N} \e^{- 2 g t} \;, \;  t \gg N^{2/3} \;,
\end{cases}
\end{eqnarray}
while in the crossover regime $t \sim {\cal O}(N^{2/3})$, $h_N(t)$ is a non-trivial random variable whose exact statistics is related to $\rho_{\rm DOS}(r,N)$ at the edge and is presently not available analytically. Again we emphasize that $h_N(t)$ is self averaging only for $t \ll {\cal O}(N^{2/3})$.

Before analyzing two-time quantities, it is already interesting to study the average value of the Lagrange multiplier $\overline{z(t)}$ which is related to the average energy
density $\overline{e(t)} = \overline{H[\{s_i(t) \}]}$. One has indeed the following relation (see also \cite{cugliandolo_dean}):
\begin{eqnarray}\label{explicit_e}
\overline{e(t)} = -\frac{\overline{z(t)}}{2} = -\frac{\overline{\lambda_{\max}}}{2} - \frac{1}{4} \overline{\frac{h_N'(t)}{{1}/{(N-1)} + h_N(t) }} \;.
\end{eqnarray}
For $t\ll {\cal O}(N^{2/3})$, one can replace $h_N(t)$ by its large $N$ value $\bar h(t)$ given in Eq. (\ref{self_av_order2}), to obtain the result given in introduction in Eq. (\ref{deltae_intro}) with the function $e_1(t)$ simply given~by
\begin{eqnarray}
e_1(t) = - \frac{1}{4} \frac{\bar h'(t)}{\bar h(t)} = 1  + \frac{1}{2t} - \frac{I_0(4t)}{I_1(4t)} \;,
\end{eqnarray}
which has the asymptotic behavior given in Eq. (\ref{energy_CD}).

Let us now analyze this formula (\ref{explicit_e}) in the limit $t \gg N^{2/3}$ where $h_N(t)$ can be replaced by its asymptotic behavior given in (\ref{asympt_hN}). One finds:
\begin{eqnarray}\label{deltae_larget}
\overline{\Delta e}(t) \sim \frac{1}{2} \overline{g \, \e^{-2gt}} \;.
\end{eqnarray}
The computation of this average requires the knowledge of the probability distribution function $p_{\rm gap}(g,N)$ of the first gap $g$, and in particular its small argument behavior since the average in Eq. (\ref{deltae_larget}) will be dominated, for large $t$, by small gaps $g$. From Ref. \cite{perret_schehr2} one has (see also \cite{WBF2013})
\begin{eqnarray}\label{asympt_gap}
p_{\rm gap}(g,N) \sim N^{2/3} \tilde p_{\rm gap}(g N^{2/3}) \;, \; \tilde p_{\rm gap}(x) \propto x^\beta \;, \; {\rm for} \; x \to 0\;,
\end{eqnarray}
for general $\beta$-ensemble. Hence, for $\beta = 1$ which corresponds to our case, one obtains that $\overline{\Delta e}(t) \propto N^{4/3}/t^3$, for $t \gg N^{2/3}$ as announced in Eq. (\ref{deltae_intro}). In the crossover regime $t\sim {\cal O}(N^{2/3})$, $\overline{\Delta e}(t)$ is a non-trivial function interpolating between the two limiting regimes (\ref{deltae_intro}), which is however hard to compute analytically.
Thus we have demonstrated that the very late time dynamics is dominated by the first gap $g$ between the two first eigenvalues of the coupling matrix. Interestingly, the same first gap also governs the fluctuations of the overlap between two spin-configurations, see \cite{MG2013}.

\subsection{Response function}

In this section, we compute the response function $R(t,t')$ defined in Eq. (\ref{def_CR}). From Eq.~(\ref{solution}) we obtain an explicit expression for such a response function as (see \ref{appendix_response} for more details)
\begin{eqnarray}
R(t,t') &=& \sqrt{\frac{g_N(t')}{g_N(t)}} \frac{1}{N} \sum_{\lambda \neq \lambda_{\max}} \e^{-(\lambda_{\max} - \lambda)(t-t')} \label{exact_R} \\
&+& \frac{1}{N} \sqrt{\frac{g_N(t')}{g_N(t)}} \left[ 1 - \dfrac{\sum_{\lambda} \e^{2 \lambda t - (\lambda_{\max} - \lambda)(t-t')}}{\sum_{\lambda} \e^{2 \lambda t}}  \right] \;, t > t' \;. \label{exact_R2}
\end{eqnarray}
 Starting from the above formula we will be able to demonstrate the existence of three different time regimes, as already mentioned in the introduction, which we analyze here in detail.

$\bullet$ Regime I where $1 \ll t' \ll N^{2/3}$ and $1 \ll t \ll N^{2/3}$. In this case, $g_N(t')$ and $g_N(t)$ are self-averaging and one can easily check that the term in Eq. (\ref{exact_R2}) is of order ${\cal O}(1/N)$ while the term in Eq. (\ref{exact_R}) is actually of order ${\cal O}(1)$. One obtains in this case \cite{cugliandolo_dean}:
\begin{eqnarray}\label{response_I}
\lim_{N\to \infty} \overline{R(t,t')} = R_{\rm I}(t,t') =  \left(\frac{t}{t'} \right)^{3/4} \bar{h}\left(\frac{t-t'}{2}\right) \;,
\end{eqnarray}
where the function $\bar{h}(x)$ is given in Eq. (\ref{self_av_order2}). As already demonstrated in Ref. \cite{cugliandolo_dean}, the dynamics for $t,t' \ll N^{2/3}$ still displays two different regimes of times, as can be seen from the structure of the response function in Eq. (\ref{response_I}). The first regime corresponds to $1\ll t,t' \ll N^{2/3}$  keeping $t-t' = \tau > 0$ fixed. In such a regime the response function is time translationally invariant, $R_{\rm I}(t,t') \sim \bar h(\tau/2)$, and so is the correlation function. In addition it was shown \cite{cugliandolo_dean} that in that regime the response and the correlation function satisfy the fluctuation-dissipation theorem (FDT). For that reason this regime is called a ``quasi-equilibrium regime''. However, it is important to note that in such a regime I the system is {\it not} at equilibrium. Indeed, taking $t,t'\gg 1$ but keeping $0<t'/t<1$ fixed one can see that the response function is clearly non-stationary. This is the so-called ``aging regime'' where, for $t \gg t'$, one has:
\begin{eqnarray}\label{r_matching_1}
R(t,t') \sim \frac{1}{4 \sqrt{2 \pi}} \frac{1}{(tt')^{3/4}} \;.
\end{eqnarray}
One can further show that in the aging regime the FDT does not hold \cite{cugliandolo_dean}.

$\bullet$ Regime II where $1 \ll t' \ll N^{2/3}$ and $t \gg N^{2/3}$. In this case $h_N(t')$ is self-averaging and given by $h_N(t') \sim 1/(4 \sqrt{2 \pi}) t'^{-3/2}$ while $h_N(t) \sim 1/N$ to leading order. On the other hand, for $t \gg N^{2/3}$, one easily sees that the second term in (\ref{exact_R}) decreases exponentially with time $t$. Finally, as $t - t' \gg N^{2/3}$ in this regime the sum over $\lambda \neq \lambda_{\max}$ in the first term in Eq. (\ref{exact_R}) is actually dominated by the first gap $g$ at the edge of the spectrum of the coupling matrix $J$:
\begin{eqnarray}
R(t,t') \sim \frac{1}{\sqrt{4 N \sqrt{2 \pi}}} \frac{1}{t'^{3/4}}  \e^{-g t} \;.
\end{eqnarray}
After averaging over the distribution of the first gap $g$ one obtains, for $t \gg N^{2/3}$, using Eq. (\ref{asympt_gap}):
\begin{eqnarray}\label{r_asympt1}
\overline{R(t,t')} \sim \frac{A_1}{\sqrt{4 \sqrt{2 \pi}}} \frac{N^{5/6}}{t'^{3/4} t^2} \;.
\end{eqnarray}
This form (\ref{r_asympt1}) is however not very illuminating. Instead, it is more convenient to rewrite $\overline{R(t,t')}$ in the scaling form valid for $ 1 \ll t' \ll N^{2/3}$ and $t \sim {\cal O}(N^{2/3})$:
\begin{eqnarray}
\overline{R(t,t')} \sim \frac{1}{t'^{3/4}} \frac{1}{\sqrt{N}} f_R\left(\frac{t}{N^{2/3}}\right) \;,
\end{eqnarray}
where $f_R(x)$ has the following asymptotic behaviors
\begin{eqnarray}
f_R(x) \propto
\begin{cases}
& x^{-3/4} \;, \; x \ll 1 \;, \\
& \\
& x^{-2} \;, x \gg 1 \;.
\end{cases}
\end{eqnarray}
Note that the small $x$ behavior indeed matches the one given in (\ref{r_matching_1}), as it should, while the large $x$ behavior is a new result. In this regime II only an aging regime is present as the time $t$ is necessarily much larger than $t'$.

$\bullet$ Regime III where both $t' \gg N^{2/3}$ and $t \gg N^{2/3}$. In this regime $h_N(t) \sim h_N(t') \sim 1/N$ while the term in (\ref{exact_R2}) decays exponentially with $t$. Hence in this regime one finds that the response function is stationary and given by
\begin{eqnarray}
\hspace*{-1.cm}\overline{R(t,t')} \sim R_{\rm III}(t-t') = \frac{1}{N} \overline{{\sum_{\lambda \neq \lambda_{\max}}} \e^{-(\lambda_{\max} - \lambda)(t-t')}}  = \int_{0}^\infty \overline{\rho_{\rm DOS}(r)} \e^{-r(t-t')}\, dr\;.
\end{eqnarray}
Hence one obtains the result announced in the introduction (\ref{rel_r_DOS}).

\subsection{Correlation function}

The correlation function $C(t,t')$ can be computed straightforwardly at zero temperature $T=0$ from Eqs. (\ref{solution}) and (\ref{def_CR}). The resulting expression is
\begin{eqnarray}\label{exact_C}
C(t,t') = \frac{g_N\left(\dfrac{t+t'}{2}\right)}{\sqrt{g_N(t) g_N(t')}} \;,
\end{eqnarray}
where the function $g_N(t)$ is given in Eq. (\ref{def_hN}). As for the response function, there are three different regimes, in the $(t,t')$ plane.

$\bullet$ Regime I where $1 \ll t' \ll N^{2/3}$ and $1 \ll t \ll N^{2/3}$: In this case, $g_N(t), g_N(t')$ and $g_N((t+t')/2)$ are self-averaging and one finds
\begin{eqnarray}
C(t,t') \sim \left( \frac{t}{t'}\right)^{-3/4} \;,
\end{eqnarray}
which is the result of Ref. \cite{cugliandolo_dean}. In this region of times the full function $C(t,t')$ is self-averaging.

$\bullet$ Regime II where $1 \ll t' \ll N^{2/3}$ and $t \gg N^{2/3}$: In this case, only $g_N(t')$ is self-averaging. By analyzing the above formula (\ref{exact_C}), one obtains the scaling form
\begin{eqnarray}
\overline{C(t,t')} \sim \left(\frac{t'}{N^{2/3}}\right)^{3/4} f_C\left(\frac{t}{N^{2/3}}\right) \;.
\end{eqnarray}
The exact computation of the full scaling function $f_C(x)$ remains a challenge but its asymptotic behaviors are  known:
\begin{eqnarray}
f_C(x) \propto
\begin{cases}
& x^{-3/4} \;, \; x \ll 1 \;, \\
& \\
& 1 \;, \; x \gg 1 \;.
\end{cases}
\end{eqnarray}

$\bullet$ Regime III where $t' \gg N^{2/3}$ and $t \gg N^{2/3}$. In this regime one has $g_N(t) \sim g_N(t') \sim g_N((t+t')/2) \sim 1/N $. We then see that to the leading order the (zero-temperature) correlation function trivializes:
\begin{eqnarray}
\overline{C(t,t')} \sim 1 \;.
\end{eqnarray}
This very simple result reflects the fact that, in regime III, $s_{\lambda}(t') \sim s_{\lambda}(t) \sim \sqrt{N} \delta_{\lambda,\lambda_{\max}}$.

\section{Conclusion}

To conclude, we have studied the zero temperature relaxational dynamics of the spherical $p=2$-spin glass model of
size $N$ when $N$ is large but finite. We have identified a crossover time scale $t_{\rm cross} \sim {\cal O}(N^{2/3})$ beyond
which the relaxation occurs within the ``well'' close to the global minimum of the energy landscape
and is dominated by saddle points with a few unstable directions. This last stage of the dynamics
is controlled by the so called density of near-extreme eigenvalues $\rho_{\rm DOS}(r,N)$ (\ref{def_DOS}) of the coupling matrix, see e.g. Eq.~(\ref{summary_rIII}), which was recently studied in the context of RMT \cite{perret_schehr,perret_schehr2}.

The problem of the long time dynamics of the same model for finite $N$ and finite temperature, $T>0$, was addressed in Ref. \cite{RM89}. The presence of finite temperature introduces a new relaxation time scale, which turns out to be the longest one in the system, related to the activation over the energy barrier separating the two ground states, that are related by spin inversion. The following simple argument proposed in \cite{RM89} shows that the energy of the minimal height of this barrier is given simply by the difference $\lambda_2-\lambda_1$. Indeed, denoting by $|\mathbf{s}_1\rangle$ (resp., $|\mathbf{s}_{2}\rangle$) the appropriately normalized eigenvector of the interaction matrix corresponding to the largest eigenvalue $\lambda_1$ (resp., second largest $\lambda_2$)  we may consider the linear combination $|\mathbf{s}_{\theta}\rangle=\cos{\theta}|\mathbf{s}_1\rangle+\sin{\theta}|\textbf{s}_2\rangle$. When the parameter $\theta$ changes continuously between $\theta=0$ and $\theta=\pi$ the vector $|\mathbf{s_{\theta}}\rangle$ describes a path along our sphere connecting  the two global minima in the associated energy landscape: $|\mathbf{s_1}\rangle$ and $-|\mathbf{s_1}\rangle$.
As easily seen the energy along such a path is given by ${E}(\theta)=-\frac{N}{2}\left(\lambda_1\cos^2{\theta}+\lambda_2\sin^2{\theta}\right)$ and is maximized at $\theta=\pi/2$, that is exactly at the saddle-point with a single unstable direction  and height given by $\lambda_2$. The barrier along this path is then obviously given by $N(\lambda_2-\lambda_1)$ and scales, in a large but finite system, like~$N^{1/3}$.

The authors of \cite{RM89} further suggested that this property will be reflected in the asymptotic behavior of the finite temperature spin-spin autocorrelation function. We hope that the approach presented here can be useful to provide a detailed analysis of such an observable for finite $N$. We leave this challenging question for future investigations.

In this paper, we restricted our study to the spherical $p=2$-spin model. Of course, it is natural to wonder whether
our results will, at least to some extent, hold for the spherical $p$-spin model with $p>2$. This is a challenging open question as, for $p>2$, the geometry of the energy landscape is much more complicated than for $p=2$, with many (exponentially with $N$) minima \cite{barriers1, barriers2} (for a rigorous proof see \cite{ABC}). Similar questions can be naturally asked about relaxation in other mean-field models with finite number $N$ of continuous degrees of freedom and exponentially many minima in the landscape, like those considered in \cite{FyoWi07,FyNa12}.

\ack

We thank L. F. Cugliandolo, J. Kurchan and G. Semerjian for useful discussions.
The research of the first author was supported by EPSRC grant EP/J002763/1 ``Insights into Disordered Landscapes via Random Matrix Theory and Statistical Mechanics''.

\appendix

\section{The questions related to self-averaging}\label{appendix_self_av}

In this appendix we study the property of being self-averaging for the density of near-extreme eigenvalues $\rho_{\rm DOS}(r,N)$. For our purpose, the concrete question is
whether $g_N(t)$, or equivalently $h_N(t)$, is self-averaging. We have studied this question in detail in the case of GUE, as we have
a full analytical description of the average DOS, $\overline{\rho_{\rm DOS}(r,N)}$, in that case. The same qualitative behavior is supposed to be valid for GOE as well.
The full computation of $\overline{\rho_{\rm DOS}(r,N)}$ was performed for the Gaussian Unitary Ensemble in Ref. \cite{perret_schehr}. From it, one can infer the scaling form of $\overline{\rho_{\rm DOS}(r,N)}$ depending on the scaling of $r$ with $N$. If $r \sim {\cal O}(1)$, $\overline{\rho_{\rm DOS}(r,N)}$ is given by a shifted Wigner semi-circle law, which was the case analyzed in \cite{cugliandolo_dean}. On the other hand, if $r \sim {\cal O}(N^{-2/3})$, one expects that $\overline{\rho_{\rm DOS}(r,N)}$ will have a different form, which is in principle hard to compute.

We have tested numerically if the following statement holds:
\begin{eqnarray}\label{self_average}
\lim_{N \to \infty} h_N(t) =  \int_0^\infty e^{-2 r t} \overline{\rho_{\rm DOS}(r,N)} dr \;,
\end{eqnarray}
where $\overline{\ldots}$ means an average over the random matrix ensemble. Our analysis shows that $h_N(t)$ is indeed self-averaging for $t \sim {\cal O}(1)$, for which the integral over $r$ is dominated by the region $r \sim {\cal O}(1)$ i.e.
\begin{eqnarray}\label{self_av_order1}
\lim_{N \to \infty} h_N(t) = \bar{h}(t) = \int_0^4 \frac{dr}{2\pi} \, e^{-2 r t} \sqrt{r(4-r)} = \frac{e^{-4t}}{2t} I_1(4t) \;, \; t \sim {\cal O}(1) \;.
\end{eqnarray}

\begin{figure}
\resizebox{100mm}{!}{\includegraphics{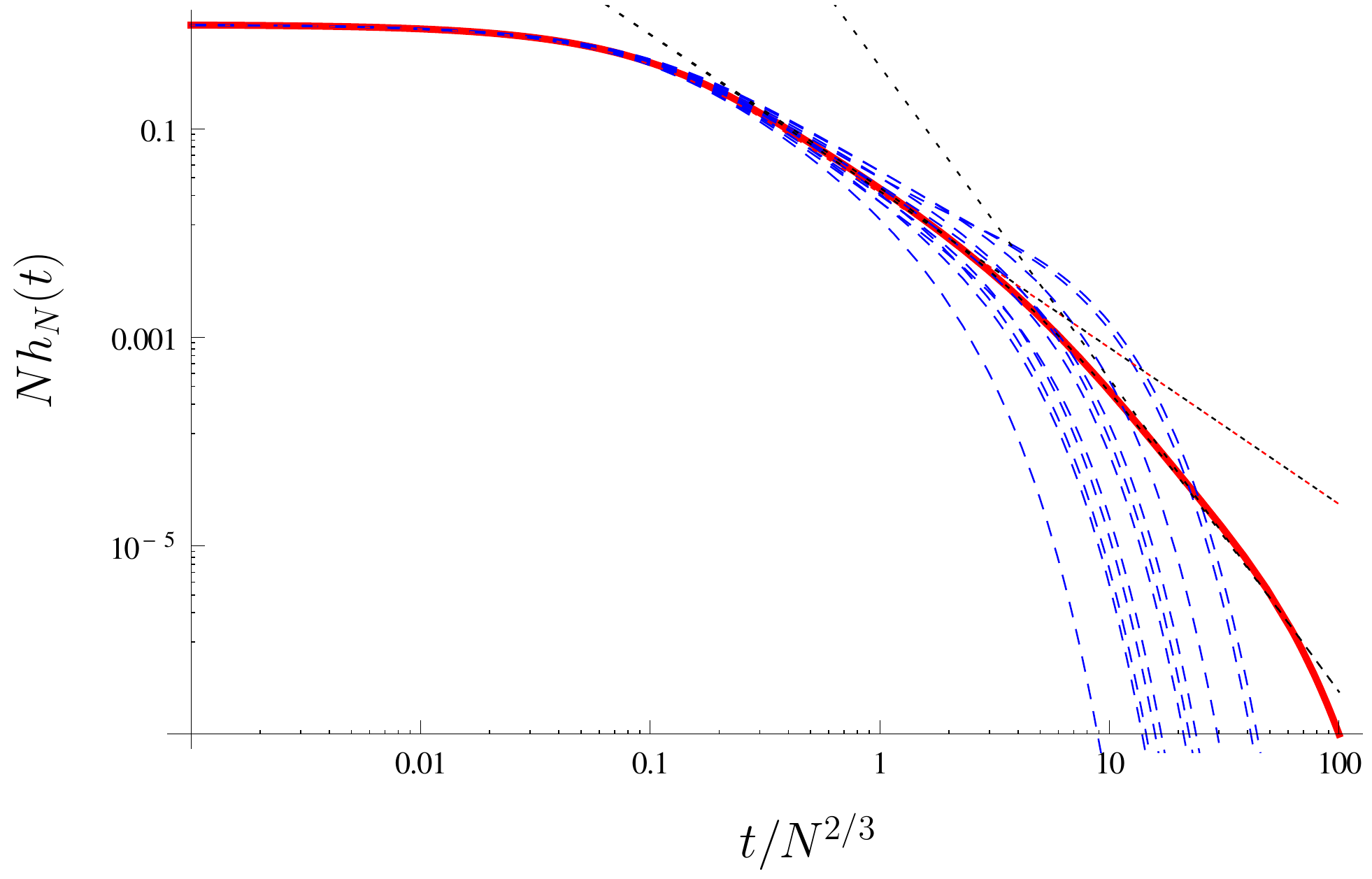}}
\caption{Plot of  $N h_{N}(t)$ as a function of $t/N^{2/3}$ for $N=50$ for GUE. In blue: 10 different simulations for $h_{N}(t)$. In red: $\overline{h_{N}(t)}$ on $10^3$ simulations. The straight lines correspond to the algebraic behaviors $t^{-3/2}$ and $t^{-1-\beta}$, with $\beta = 2$ here for GUE.
}\label{fig_self_average}.
\end{figure}

As we show now, self-averaging, as stated in (\ref{self_average}) does not hold when $t \sim {\cal O}(N^{2/3})$, i.e. when the integral over $r$ is  dominated by $r \sim {\cal O}(N^{-2/3})$. Indeed, if self-averaging (\ref{self_average}) held we would have
\begin{eqnarray}\label{expected}
h_N(t) \sim \frac{1}{N} \tilde h\left(\frac{t}{N^{2/3}} \right) \;, \; \tilde h(x) = \int_0^\infty e^{-2 x u} \tilde \rho_{\rm edge}(u) du \;.
\end{eqnarray}
In Fig. \ref{fig_self_average}, we show a plot of $N h_N(t)$ as a function of $t/N^{2/3}$ for 10 different realizations of $N \times N$ GUE
random matrices, with $N=50$. We see clearly that the different curves do coincide for $t/N^{2/3} \ll 1$ -- corresponding to the bulk regime (\ref{expected})
but these curves clearly differ as soon as $t/N^{2/3} \sim 1$, which is a clear indication that self-averaging does not hold in this regime.

\section{Computation of the response function $R(t,t')$}\label{appendix_response}

In this appendix, we give the details of the computation leading to the expression for the response function $R(t,t')$ given in Eq. (\ref{exact_R2}). The starting point of this computation is the expression for the response given in Eq. (\ref{def_CR}), for $t>t'$:
\begin{eqnarray}
R(t,t') = \frac{1}{N} \sum_{\lambda} \frac{\delta s_{\lambda}(t)}{\delta h_\lambda(t')} \Big|_{h=0} \;,
\end{eqnarray}
where $s_\lambda(t)$ evolves via the Langevin Eq. (\ref{def_Langevin}) with $\xi_\lambda(t) = 0$, for all $\lambda$ and $t$ (since we are interested here in the zero temperature $T=0$ dynamics):
\begin{eqnarray}\label{def_Langevin_app}
\frac{\partial s_\lambda(t)}{\partial t} = (\lambda - z(t)) s_\lambda(t) + h_\lambda(t) \;,
\end{eqnarray}
starting from the initial condition $s_\lambda(t=0) = 1$ for all $\lambda$. Here $h_\lambda(t)$ represents an (infinitesimal) external magnetic field and $z(t)$ is a Lagrange multiplier imposing the spherical constraint $\sum_\lambda s_\lambda^2(t) = N$. Hence $z(t)$ depends on $h_\lambda(t')$ for all $\lambda$ and $0<t'\leq t$. The solution of this Langevin equation is given in Eq. (\ref{solution}):
\begin{eqnarray}\label{solution_appendix}
&&s_\lambda(t) = \exp(\lambda t) \exp{\left[-\int_0^t z(\tau) d\tau \right]} \\ \nonumber
&&+ \int_0^t dt'' \exp{[\lambda(t-t'')]} \exp{\left[- \int_{t''}^t z(\tau') d\tau' \right]\,h_\lambda(t'')} \;.
\end{eqnarray}
From this expression we see that $s_\lambda(t)$ depends both {\it explicitly} on $h_\lambda(t)$ (see the second term) and {\it implicitly} through the Lagrange multiplier~$z(t)$. It is convenient to introduce the function
\begin{eqnarray}\label{def_gnh}
g_{N,h}(t) = \exp{\left[ 2\int_0^t dt'(z(t') - \lambda_{\max}) \right]} \;,
\end{eqnarray}
such that $g_{N,h=0}(t) = g_N(t)$ given in Eq. (\ref{expr_z_2}). In terms of $g_{N,h}(t)$ on has, from~Eq.~(\ref{solution_appendix}):
\begin{eqnarray}\label{eq:derivative_1}
\hspace*{-1cm}\frac{\delta s_{\lambda}(t)}{\delta h_\lambda(t')} \Big|_{h=0} = \sqrt{\frac{g_N(t')}{g_N(t)}}\, \e^{(\lambda-\lambda_{\max})(t-t')} - \frac{1}{2 \, [g_N(t)]^{3/2}} \e^{(\lambda-\lambda_{\max})t} \frac{\delta g_{N,h}(t)}{\delta h_\lambda(t')} \Big|_{h=0} \;.
\end{eqnarray}
The term $\frac{\delta g_{N,h}(t)}{\delta h_\lambda(t')} \Big|_{h=0}$ in Eq. (\ref{eq:derivative_1}) can be computed from the normalization condition $\sum_\lambda s_\lambda^2(t) = N$. Indeed, by expanding the latter relation, using Eq. (\ref{solution_appendix}), to lowest order in $h_\lambda$'s one obtains
\begin{eqnarray}\label{expansion}
&&\hspace*{-2cm}\sum_{\lambda} \Bigg[\e^{2 \lambda t - 2 \int_0^t d\tau z(\tau)} + 2  \e^{\lambda t - \int_0^t d\tau z(\tau)} \int_0^t dt'' \exp{[\lambda(t-t'')]} \exp{\left[- \int_{t''}^t z(\tau') d\tau' \right]\,h_\lambda(t'')} \nonumber \\
&& +Ê{\cal O}(h_\lambda^2) \Bigg] = N \;.
\end{eqnarray}
By differentiating this relation (\ref{expansion}) with respect to $h_\lambda(t')$ and setting $h_\lambda(t'') = 0$ for all $0< t''\leq t$, one finds [using the definition of $g_{N,h}(t)$ in Eq. (\ref{def_gnh})], from the explicit solution in Eq. (\ref{solution_appendix}):
\begin{eqnarray}\label{eq:derivative_2}
\hspace*{-2cm}-\frac{1}{[g_N(t)]^2} \frac{\delta g_{N,h}(t)}{\delta h_\lambda(t')} \Big|_{h=0} \sum_{\lambda} \e^{2(\lambda-\lambda_{\max})t} + 2 \e^{-(\lambda_{\max}-\lambda)t - (\lambda_{\max}-\lambda)(t-t')} \frac{\sqrt{g_N(t')}}{g_N(t)} = 0 \;.
\end{eqnarray}
Therefore one has
\begin{eqnarray}\label{deltag}
\hspace*{-1cm}
\frac{\delta g_{N,h}(t)}{\delta h_\lambda(t')} \Big|_{h=0} = \frac{2}{\sum_{\lambda} \e^{-2(\lambda_{\max}-\lambda)t}} \e^{-(\lambda_{\max}-\lambda)t - (\lambda_{\max}-\lambda)(t-t')} g_N(t) \sqrt{g_N(t')} \;.
\end{eqnarray}
By injecting this relation (\ref{deltag}) into Eq. (\ref{eq:derivative_1}) one arrives at
\begin{eqnarray}
\hspace*{-1.5cm}R(t,t') = \sqrt{\frac{g_N(t')}{g_N(t)}} \frac{1}{N} \sum_\lambda \e^{-(\lambda_{\max} - \lambda)(t-t')} - \frac{1}{N} \sqrt{\frac{g_N(t')}{g_N(t)}}  \dfrac{\sum_{\lambda} \e^{2 \lambda t - (\lambda_{\max} - \lambda)(t-t')}}{\sum_{\lambda} \e^{2 \lambda t}}  \;,
\end{eqnarray}
which yields (by isolating the term $\lambda = \lambda_{\max}$ in the first sum) the expression given in the text in Eq. (\ref{exact_R2}).

\end{document}